\documentclass[preprint]{aastex}

\tighten
\slugcomment{Accepted for publication in 
{\it The Astrophysical Journal}}

\begin{document}


\newcommand{\etal}{{et~al.}}
\newcommand{\noi}{\noindent}
\newcommand{\bs}{\bigskip}
\newcommand{\ms}{\medskip}
\renewcommand{\ss}{\smallskip}
\renewcommand{\deg}{$^{\circ}$}
\newcommand{\um}{$\mu$m}
\newcommand{\uk}{$\mu$K}
\newcommand{\qrms}{$Q_{rms-PS}$}
\newcommand{\n}{$n$}
\newcommand{\gt}{$>$}
\newcommand{\lt}{$<$}
\newcommand{\il}[1]{$I^3_{#1}$}
\newcommand{\ing}{$I^3_{16}$}
\newcommand{\iell}{$I^3_{\ell}$}
\newcommand{\cobe}{$COBE$-DMR}


\shorttitle{Non-Gaussianity in the \cobe\ Sky ?}
\shortauthors{Banday, Zaroubi \& G\'orski}

\title{On the Non-Gaussianity observed in the \cobe\ Sky 
Maps}

\author{A.J. Banday\altaffilmark{1} and S. Zaroubi\altaffilmark{2}}
\affil{Max Planck Institut f\"{u}r Astrophysik, D-85740 Garching 
bei M\"unchen, Germany.}

\and

\author{K.M. G\'orski\altaffilmark{3,4,5}}
\affil{Theoretical Astrophysics Center, Juliane Maries Vej 30,
DK-2100, Copenhagen 0, Denmark.}

\altaffiltext{1}{{\it email:} banday@mpa-garching.mpg.de}
\altaffiltext{2}{{\it email:} saleem@mpa-garching.mpg.de}
\altaffiltext{3}{Warsaw University Observatory, 
                 Aleje Ujazdowskie 4, 00-478 Warszawa, Poland.}
\altaffiltext{4}{Current address: ESO, D-85740 Garching 
bei M\"unchen, Germany.}
\altaffiltext{5}{{\it email:} kgorski@eso.org}


\begin{abstract}

In this paper we pursue the origin of the non-Gaussianity determined
by a bispectrum analysis of the \cobe\ 4-year sky maps. The
robustness of the statistic is demonstrated by the rebinning of
the data into 12 coordinate systems. By computing the bispectrum statistic
as a function of various data partitions -- by channel, frequency, and
time interval, we show that the observed non-Gaussian signal is driven
by the 53 GHz data. This frequency dependence strongly rejects the 
hypothesis that the signal is cosmological in origin. 
A jack-knife analysis of the coadded 53 
and 90 GHz sky maps reveals those sky pixels to which the bispectrum statistic
is particularly sensitive. We find that by removing data from the 53
GHz sky maps for periods of time during which a known systematic effect
perturbs the 31 GHz channels, the amplitudes of the bispectrum coefficients
become completely consistent with that expected for a Gaussian sky.
We conclude that the non-Gaussian signal detected by the
normalised bispectrum statistic in the publicly available DMR sky maps
is due to a systematic artifact.
The impact of removing the affected data on estimates
of the normalisation of simple models of cosmological anisotropy
is negligible.

\end{abstract}


\section{Introduction}

 Gaussianity plays a fundamental role in modern physics. For a
Gaussian field, the power spectrum tells us all there
is to know about that field. In cosmology too, the role of Gaussian
statistics is of great importance.  Inflation (or at least the
simplest models thereof) produces density
perturbations which are random-phase
and have amplitudes with a Gaussian distribution on any given scale. 
Since the onset of non-linear evolution for the density
perturbations can itself create non-Gaussian correlations, it is simplest
to test the statistics of the perturbations in the linear regime. 
Such scales are cleanly probed by anisotropies in the Cosmic Microwave
Background (CMB) as imprinted by density perturbations on the last
scattering surface.

Since 1992 (Smoot \etal, 1992) the \cobe\ maps of the microwave sky 
at 31.5, 53 and 90 GHz have been proferred as strong evidence in support
of the inflationary paradigm. Many tests applied to the data have 
shown consistency with Gaussianity. 
For example, skewness and kurtosis (Smoot \etal, 1994),
the three-point correlation function (Hinshaw \etal, 1994, 1995; 
Luo, 1994), the
correlation function of temperature extrema (Kogut \etal, 1995),
the genus (Smoot \etal, 1994; Colley, Gott \& Park, 1996), 
a compendium of the above
as applied to the 4-year sky maps (Kogut \etal, 1996a),
the bispectrum (Heavens, 1998) and Minkowski functionals (Schmalzing
\& G\'orski, 1998). Recent analyses
by Pando, Valls-Gabaud \& Fang (1998) using wavelets and Ferreira,
Magueijo \& G\'orski (1998) utilising a normalised bispectrum
estimator find evidence for non-Gaussianity at levels of 98\% or
above\footnote{
An additional paper by Novikov, Feldman \& Shandarin (1999)
applying partial Minkowski Functionals to the \cobe\ DCMB and DSMB 
foreground removed sky maps has been cited by some as additional 
evidence for non-Gaussianity.
However, they apply their analysis to the full sky which is simply
incorrect given the known inadequacies of the foreground removal
techniques. This results in several obvious residual artifacts,
especially in the Galactic Center region. We point out that
the authors themselves do not consider that their analysis has detected 
any significant non-Gaussianity.}.
Such a result is simply remarkable:
even the major theoretical competitor to inflation -- models which invoke
networks of cosmological defects to seed structure formation -- only show
evidence of non-Gaussian CMB behaviour in subtle ways. 
In fact, one still expects the large-scale anisotropy seen by \cobe\
for a typical defect model 
to appear quite Gaussian (as a consequence of the typical coherence
scale of a defect being small relative to the DMR beam and an
application of the central limit theorem). The detection of
{\it cosmological} non-Gaussianity in the DMR data would have profound
implications for all models of structure formation.

Before moving on to our own analysis, we would like to point out that 
we do not consider it correct to \lq de-weight' the significance of
the new results given that earlier tests failed to reject Gaussianity. 
In particular, non-Gaussianity has no generic signature on the sky,
and as such different tests will demonstrate variable statistical
power depending on the type of non-Gaussian feature being sought. 
For any statistic, there are relative merits and disadvantages 
depending on whether the non-Gaussian features are localised in Fourier
or real space, the effective scales probed by the statistic, and the
impact of instrumental noise on the analysis (as a simple case, the
skewness was shown to be biased by noise in Smoot \etal, 1994). 
It is certainly possible that a model can appear Gaussian over many
scales but then suddenly demonstrate very non-Gaussian behaviour.

In what follows, we will attempt to determine the origin of the
non-Gaussian signal detected in the \cobe\ 4-year sky maps
using the bispectrum technique of Ferreira, Magueijo \& G\'orski 
(1998 -- hereafter FMG). After reviewing the statistic itself and
summarising the current status of the bispectrum analysis, we
will determine the stability of the computed bispectrum amplitudes
as the data is rebinned into 12 different coordinate systems, 
investigate the presence of signal in various data partitions 
by channel and year, infer a region of the DMR sky to which the
statistic is particularly sensitive, and finally propose a plausible
origin of the detected signal based on a known systematic error.

\section{The Bispectrum Statistic}

In this section we consider the definition of a {\em normalised
bispectrum} as defined in FMG. In the spherical harmonic
representation one can expand the temperature field on the celestial
sphere in terms of $a_{\ell m}$ coefficients,
\begin{equation}
{\Delta T}({\bf n})=\sum_{\ell m}a_{\ell m}Y_{\ell m}({\bf n})
\end{equation}

If we consider a multipole coefficient of the field
\begin{equation}
\Delta T_\ell=\sum_{m}a_{\ell m}
\end{equation}
then by forming the tensor product of 3 such coefficients and
requiring rotational invariance one computes an estimator of 
the bispectrum as
\begin{eqnarray}
{\hat B}_\ell&=&\alpha_\ell\sum_{m_1m_2m_3}
{\cal W}^{\ell \ell\ell}_{m_1m_2m_3} a_{\ell m_1}a_{\ell m_2} a_{\ell m_3}
\nonumber \\
\alpha_\ell&=&\frac{1}{(2\ell+1)^{\frac{3}{2}}}\left (
{\cal W}^{\ell \ell\ell}_{0 0 0} \right )^{-1}
\label{bispec}
\end{eqnarray}
where ${\cal W}^{\ell_1\ell_2\ell_3}_{m_1m_2m_3}$ are the Wigner 3-J
coefficients (see, for example, Messiah 1976). 

Finally, FMG require a statistic which is invariant under a parity 
transformation and dimensionless. They define $I^3_\ell$ to be
\begin{eqnarray}
I^3_\ell &=&\left| { {\hat B}_\ell\over ({\hat C}_\ell)^{3/2}}
\right| \label{defI}
\end{eqnarray}
where ${\hat C}_\ell=\frac{1}{2\ell+1}\sum_m|a_{\ell m}|^2$ (ie. the
power spectrum). 
This is the normalised bispectrum we refer to in what follows.
Only EVEN values of $\ell$ lead to non-zero
bispectrum amplitudes due to the properties of the Wigner 3-J
coefficients. 

The bispectrum coefficients are evaluated from an input sky map as
follows. We apply the extended Galactic cut of Banday \etal\ (1997)
to the sky map in order to excise those pixels with significant
Galactic contamination near the plane of the Galaxy, subtract the
best-fit monopole and dipole (as computed on the cut-sky)
from the remaining pixels\footnote{Note 
that this differs slightly from the procedure
of FMG where only the monopole is subtracted. We have checked that
we recover the FMG bispectrum values 
when we also omit the dipole subtraction. }, 
then compute the harmonic amplitudes $a_{\ell m}$ according to eqn (1).
(3) and (4) subsequently allow the bispectrum values to be determined.
Monte Carlo simulations of the sky as seen by \cobe\ are then
used to compute the statistical distributions of the various 
\iell\ modes, since these are  themselves inherently non-Gaussian.
The distributions are then compared to the data in order to assess
statistical significance. 

\section{Summary of the Current Situation}

The above method has been applied to the coadded 53 and 90 GHz
\cobe\ 4-year sky maps by FMG. Fig.~1 shows our results 
(for Galactic frame data)
to be compared with the corresponding Figure in their paper.
Clearly, we reproduce their results to high accuracy. The main
features of the FMG results are as follows: 
based on a \lq\lq $\chi^2$''-like analysis of the bispectrum results
FMG find that the DMR coadded 53 and 90 GHz data are inconsistent
with Gaussianity at more than 98\% c.l.;
that this result is driven by the $I^3_{16}$ mode; and that
the bispectrum amplitudes are noise dominated beyond $\ell$ = 18.

Possible objections to the statistical significance of the
analysis or suggestions as to plausible non-cosmological 
sources of the non-Gaussian signal have been addressed
in several papers by the same authors (Magueijo, Ferreira \&
G\'orski, 1998, 1999; Ferreira, G\'orski \& Magueijo, 1999).
To summarise:
\begin{itemize}
  \item the result is not sensitive to assumptions about the cosmological
  model, noise properties of the data, Galactic cut or other details
  of the analysis
  \item systematic error templates show no comparable signal
  \item correcting for Galactic foreground contamination in the usual
  way actually increases the confidence level for rejection of
  Gaussianity to over 99\%
\end{itemize}
The latter result is quite intriguing in itself. Intuitively, one
might feel that the Galactic foregrounds are more likely to
demonstrate non-Gaussian behaviour than the CMB, yet this does not
appear to be the case. Presumably, any non-Gaussianity manifests
itself on scales other than those measured here, or, as with
cosmological defect models, in more subtle ways. More importantly,
the increase in the confidence level for rejection of Gaussianity
answers those skeptics who might consider the original
detection to be unconvincing. Given the previously held belief
of the pristine nature of the \cobe\ data, we proceed to expand
on the FMG analyses.

\section{Stability of the Bispectrum Statistic}

The issue of the stability of the bispectrum estimator is of paramount
importance to the legitimacy of the detection of non-Gaussianity.
The microwave data obtained by the \cobe\ instruments are 
subjected to an initial form of data reduction, namely the
construction of full-sky maps. This is an important step, since it
compresses the information contained in a large data-stream into a 
more managable size. At this map-making step, various systematic error
corrections are applied, and the data is essentially binned into
pixels with a preferred orientation and geometry on the sky, depending
on both the pixelisation scheme and coordinate reference frame
adopted. The reader should consult White 
\& Stemwedel (1992) for a description of the $COBE$ Quadrilaterized 
Spherical Cube -- hereafter quadcube -- pixelisation scheme.
Obviously, one attempts to choose a pixel size which 
samples the sky signal in a lossless fashion.  It is certainly the case,
however, that certain systematic and/or noise signals may add
more coherently in particular coordinate frames.
This is a consequence of the result that there is no unique one-to-one 
mapping of pixels from one frame to another, since the pixels are
of finite size and themselves have specific geometries.
In essence, the signal and noise 
are rebinned from one frame to another, but one would naturally 
expect a robust statistical estimator to render results essentially 
independently of the coordinate frame. FMG recognise this important 
issue and do compare results in the publicly available Galactic and Ecliptic 
frames showing them to be very consistent. We extend their analysis
further by analysing an additional 10 coordinate frames.

This important test will also allow us to comprehensively assess the
impact of two other aspects of the analysis. The most obvious
feature in the full-sky maps is the Galactic plane emission which must
be discarded from the analysis. Some small \lq leakage' of the strong Galactic
emission into pixels close to the plane will vary slightly 
in the different coordinate schemes thereby testing whether the
detected signal is due to residual Galactic plane signal
surviving the cut. More importantly, as a consequence of the
necessity to excise pixels from the analysis, the $a_{\ell m}$ 
coefficients we substitute in equation (4) contain aliased signal
due to the incomplete sky coverage. The geometry of the Galactic plane
cut in a given coordinate system determines the relative aliases contributions
from other $\ell$ modes. Thus we can determine whether the detected
signal is a chance aliasing of signals from other modes
\footnote{Of course, this is one of the reasons why Monte Carlo
simulations are used to assess the significance of the bispectrum 
amplitudes. These should account for the aliasing issue at least 
statistically.}.

Fig.~2 shows our 12 reference frames. 4 correspond to conventional
coordinate schemes of astronomical relevence -- Galactic, Ecliptic,
Celestial and Supergalactic. The remaining 8 were designed
to place the Galactic plane in specific orientations with respect to
the quadcube.
Whilst there are an infinite number of ways in which to rotate and
re-pixelise the data, the quadcube faces and edges provide 
obvious axes along which to align the Galactic plane and test the
issues discussed in the previous paragraph. In particular, we evaluate
whether there 
exists a conspiracy between the pixelisation scheme and the effect of data
omission in the bispectrum analysis, the results of which follow.

In Fig.~3 we present the results of our analysis for the coadded 53
and 90 GHz data. 
There is excellent consistency of the \iell\ values
for all coordinate frames, with the peak-to-peak 
scatter between frames less than
$\sim$ 0.3 per mode. This explicitly establishes the rotational 
invariance of the estimator.
The \ing\ mode has a well-defined mean value 
inconsistent with Gaussian behaviour. We note, however, that there
are coordinate frames where this observed bispectrum amplitude
is less significant as evidence for non-Gaussian behaviour,
but also those where the reverse is true\footnote{
In order to assess the significance of any differences observed, 
we have performed simulations to demonstrate that the
distribution of \iell\ values is essentially independent of the 
coordinate frame. Thus we can simply compare bispectrum amplitudes
computed in different frames.}. 
This might be interpreted
by some as indicating a lack of robustness in the statistic. We
believe that the observed mean value is significant, and that the
scatter can be due to some variation in aliasing in the
different schemes. Simply put, the non-Gaussian signal itself is actually
suppressed by an increased aliased contribution from Gaussian modes
in some cases. The behaviour of the \ing\ mode
is clearly very different from the others and indicative of a genuine
non-Gaussian feature in the coadded sky map. 
While the values for the $\ell$=6 and 14 modes also appear quite high, they
lie comfortably within the probability distributions summarised in
Fig.~1.


\section{Channel Dependence of the Bispectrum Signal}

In this section we initially attempt to isolate the observed 
bispectrum signature to a particular channel or frequency. 
One should keep in mind that the signal-to-noise of the statistic
varies as (Number of Observations)$^{\frac{3}{2}}$, so that
such attempts are hindered by lower sensitivity. Still, we do
find some revealing properties of the data. In the remainder
of the section, we concentrate on the \ing\ mode predominantly.

Fig.~4 shows a compilation of results for the 53 and 90 GHz
channels. The most important feature is the large \ing\ amplitude
associated with the 53 GHz signal maps. For the sum map, although 
there are variations from frame-to-frame, these are 
consistent with those seen for the coadded data, despite
lower signal-to-noise. For the individual channels, the situation
is less clear, particularly for the 53B sky maps where in no case
would one claim a detection of non-Gaussianity. The 53A does
seem to contain strong hints of such behaviour. The 90 GHz channels,
however, show no unusual properties whatsoever. The most appealing
inference to make, therefore, is that the bispectrum signal 
in the coadded sky map is dominated by the contribution from the
53 GHz channels. If this is indeed the case, then we are forced
to conclude that the underlying non-Gaussian signal, whatever its
origin, is non-cosmological. It is also unlikely that the 
signal is noise-related since the difference maps show little
evidence for significant bispectrum amplitudes.

Despite the decreasing returns expected, we consider 
the Galactic sky maps generated by partitioning the data
into yearly or two yearly sections to determine whether
we can isolate the cause of the detected bispectrum signal
to a particular time period. This analysis proved less revealing:
there is little strong evidence of a noteworthy signal
present in any maps for \ing\ (Fig.~5). In fact, many $\ell$-modes show a 
large scatter from year-to-year. In an Appendix, we demonstrate 
via simulations that this is reasonable for such noise-dominated sky maps.

\section{Does the Bispectrum Signal have a  Spatial Signature?}

Having asserted that the observed \ing\ signal is principally
associated with a single frequency, we now attempt to determine
whether the effect can be isolated spatially. In other words,
can we identify specific pixels or groups of pixels to which our
statistic is sensitive? We apply a \lq jack-knife' type analysis
to the maps: for every pixel surviving the Galactic cut, we recompute
the bispectrum amplitude after removing that pixel and its 4 nearest
neighbours. To visualise this, we plot as a sky map the
bispectrum amplitude such that the value at a given pixel corresponds
to the \iell\ amplitude on removal of that pixel and its neighbours.
Fig.~6a shows results for the interesting $\ell$ = 16
mode in three coordinate systems -- Galactic, Ecliptic and
System 1. The general pattern (which is consistent upon rotation of
the various frames to Galactic) reflects the underlying 
$\ell$ = 16 mode. Nevertheless, there are clusters of pixels
to which the bispectrum shows enhanced sensitivity. In particular,
there are standout features in the Northern hemisphere (located in
Galactic coordinates on Face 0 of the quadcube). 
One should recall that this is the region where Pando \etal\ (1998)
detect a non-Gaussian signal using a wavelet method, and also that 
Magueijo \etal\ (1999) suggest a localisation of the non-Gaussian
signal to the Northern Galactic hemisphere.
Fig.~6b shows the \ing\ amplitude
directly as a function of removed pixel index. Again, we note that 
System 1 has a lower average value for the bispectrum amplitude.
However, for each system, the mean amplitude is well defined
with typically small scatter\footnote{
The statistical means and variances for the Galactic, Ecliptic and
System 1 frames are, respectively, $0.939\pm\ 0.043$, $0.912\pm\ 0.036$ 
and $0.714\pm\ 0.040$.}, but a few very large excursions which increase
or reduce the bispectrum value dramatically. 
Other bispectrum values show similar global behaviour related to their
progenitor $\ell$ modes, but with notably less sensitivity to
localised pixel clusters.
Interestingly, Bromley \& Tegmark (1999) have suggested that removing
a pixel centred on $b$ = 39.5\deg\ and $l$ = 257\deg\ (pixel 4845 in
Galactic coordinates) plus its 4 nearest neighbours reduces the \ing\
bispectrum amplitude to a statistically insignificant level\footnote{
In our analysis, this procedure reduces the observed bispectrum 
amplitude from 0.949 to 0.678.}, and
propose that this reflects on the robustness of the bispectrum
estimator. Our results show indeed that there are pixels to which the
bispectrum is sensitive, but that the amplitude can be both
significantly enhanced as well as reduced. This implies that there is
a real spatial distribution of pixels which is the cause of the
interesting \ing\ signal. 

As an additional remark, we respond to claims 
that the observed signal may show dependence to localised noise. 
We have repeated our analyses on the sky maps after \lq cleaning' 
them, ie. after removing all power above $\ell$ = 40 (which
is almost entirely noise dominated) as computed on the
full-sky. There is no significant change in any of our results
after performing this procedure, which, combined with the spatial signature
noted above, refutes at least in part this suggestion.

\section{Impact of a Known Systematic Effect}

As a result of our jack-knife analysis, we have found tentative
evidence for a distribution of pixels on the sky which the data
is sensitive to. We now attempt to attribute this signal 
with a systematic effect present in the time-ordered-data (TOD).
Since the pixels affected do not fully cover the sky,
any potential contaminant in the TOD should have a periodicity less 
than the six-months required to map the celestial sphere completely.

The \lq eclipse effect' is an orbitally modulated signal taking place 
for approximately two months every year around the June solstice
when the $COBE$ spacecraft repeatedly flies through the
Earth's shadow.
The 31 GHz channels are particularly adversely affected, with obvious
trends in the TOD present (see Fig.~2 of Kogut \etal, 1992) which 
are strongly correlated with several
housekeeping signals relating to the spacecraft thermal and electrical
properties. However, an attempt to model and correct the erroneous
signal based on this correlation was only partially successful: 
the model removes only two-thirds of the anomalous signal
in the 31GHz channels. Consequently, the 31 GHz sky maps do not include
data taken during the eclipse season. Only very small effects were 
noted for the
53 and 90 GHz channels, which were therefore corrected by the empirical
model. Kogut \etal\ (1996b) estimated that only 0.3 \uk\ rms artifacts 
are introduced into these channels if the eclipse data is included.
Nevertheless, there are several hints that the eclipse period may
be responsible for the observed bispectrum signal. 

Firstly, the 31 GHz channels in which the eclipse data are rejected are
extremely well-behaved with respect to the bispectrum tests,
yet when we perform a similar analysis of the 31A channel 
with the eclipse data included one observes a notable enhancement 
of the $I^3_{16}$ mode. This is quite suggestive despite the
observed amplitudes being comfortably Gaussian in both cases, and the
signal-to-noise being quite low (Fig.~7). 

Secondly, the region of sky predominantly affected by removing 
the eclipse data aligns quite well with those pixels to
which the bispectrum is apparently sensitive. Fig.~8 shows the 
temperature difference between the coadded 53 and 90 GHz
sky maps made both with and without the eclipse data.

We therefore proceed to remove the eclipse periods from the 53 and 90
GHz data and study its impact on the bispectrum calculation. Fig.~9
summarises the situation. We note that:
\begin{itemize}
  \item there is a remarkable drop in the $I^3_{16}$ mode for the
  53A channel, so that it is now practically consistent with zero, ie
  very Gaussian, and similarly an important drop in the 53B channel
  \item the 90 GHz channels remain largely unaffected and therefore 
  comfortably consistent with Gaussianity
  \item the coadded 53 and 90 GHz sky map is now fully consistent with
  Gaussianity.
\end{itemize}
The impact on the 53A channel is particularly dramatic, the crucial
issue being whether such a change in the statistic's value can be more
simply attributed to the no-eclipse maps being noisier. We have
performed a large number of simulations of CMB skies with noise 
properties corresponding to the two situations here. There are almost no
cases where such a precipitous drop in the $I^3_{16}$ occurs: 
in 20000 simulations we find only two cases where \ing\ drops by $\sim$
0.8 when comparing the no-eclipse with eclipse simulations. In fact,
a mere 195 simulations demonstrate a drop of 0.5.
This should not seem surprising: in our previous analysis of the 
changes of $I^3_{16}$ from a yearly map to 4-year sky map, 
we see no similar behaviour for any mode, although the change in the noise
contribution is, in this case, much greater.

We further note that a
coadded sky map constructed from 53 GHz channels excluding the eclipse
months and 90 GHz data including this data also demonstrates
a drop in the bispectrum amplitude, whereas the reverse combination
does not. This supports our earlier claim that
the non-Gaussian signal is strongly associated with a single frequency
and thus non-cosmological.

Since FMG found that the bispectrum amplitude at $\ell$ = 16
was enhanced after subtraction of the Galactic foreground model,
we repeat our analysis after correcting the no-eclipse coadded
sky map. Although once again, there is some small increase in the
amplitude of the formerly spurious mode, it still remains 
perfectly consistent with the expectations of a Gaussian 
model for the CMB signal and instrumental noise\footnote{
The coadded (53A) sky map has an \ing\ value of 0.949 (0.791)
which falls to 0.522 (0.012) on excluding the eclipse data. 
If we correct the data with the best fitting Galactic foreground model, the
corresponding numbers are 1.112 (0.840) falling to 0.527 (0.116).
.}

\section{Impact on Previous Cosmological Results}

The impact of the data excision on other analyses of the DMR
data, especially the power spectrum fits to cosmological models,
should be investigated. Although a comprehensive reassessment of
all of the models previously constrained by the \cobe\ data is
beyond the scope of this paper, we have recomputed fits to a pure
power law model of initial perturbations as parameterised by
\qrms\ and \n. For an analysis of the coadded 53 and 90 GHz data
in Galactic coordinates including the eclipse data, we find
\[
  \mathrm{Q_{rms-PS}}\ =\ 15.63^{+3.20}_{-2.56}\ \mu \mathrm{K},\\
  \mathrm{n}\ =\ 1.22^{+0.23}_{-0.27}
\]
\[
  \mathrm{Q_{rms-PS}\mid_{n=1}}\ =\  18.44^{+1.3}_{-1.2}\ \mu \mathrm{K}
\]

If we exclude the eclipse data
\[
  \mathrm{Q_{rms-PS}}\ =\ 16.27^{+3.20}_{-2.88}\ \mu \mathrm{K},\\
  \mathrm{n}\ =\ 1.15^{+0.29}_{-0.25}
\]
\[
  \mathrm{Q_{rms-PS}\mid_{n=1}}\ =\ 18.25^{+1.4}_{-1.3}\ \mu \mathrm{K}
\]
These results are summarised in Fig.~10. The change in the best fit
model is consistent with an increase in the average noise level
per pixel of $\sim$ 10\%, and is comparable with the difference
in fits between Galactic and Ecliptic frame data (cf. Table 1 of
G\'orski \etal, 1996). We consider that it is unlikely that fits
to other cosmological models will be affected with any greater
significance.

It will, of course, be interesting to determine whether the eclipse
effect is responsible for the wavelet detection of non-Gaussianity,
especially given that the effect is localised on Face 0 of the 
$COBE$ quadcube. We defer this to later work.

\section{Summary}

We believe that we have proposed a convincing non-cosmological 
candidate responsible for the observed non-Gaussian signal
in the \cobe\ 4-year sky maps.. 
That this candidate is, in fact, a known systematic artifact 
in the data is not unreasonable. Although the 31 GHz channels
were affected by the eclipse data in such an intrusive way
that the exclusion of the data was warranted, its impact on the
53 GHz channels is more subtle, and only revealed by a new statistic
sensitive to non-Gaussian signals in the data. Presumably, the 
source of the signal corresponds to that part of the eclipse effect
that could not be removed from the 31 GHz channels. In any case,
excluding the data from the 53 GHz channels removes the non-Gaussian
behaviour. We conclude, therefore, that the non-Gaussianity present in the
publicly released DMR 4-year sky maps is not cosmological in origin. 
Fortunately,
it appears that for most cosmological issues, the presence of this
artifact is of no importance: analysis of the coadded sky maps 
excluding the eclipse data provides fits to power-law 
cosmological models completely consistent with previous work.

\acknowledgements
We thank Gary Hinshaw for useful suggestions.
We acknowledge the efforts of those contributing to the \cobe\
and particularly recognise the map-making skills of Phil Keegstra.

\appendix
\section{Noise dependence of the Bispectrum Statistic}

An important issue in our statistical analysis of the
bispectrum values is the dependence of the 
statistic on the signal-to-noise ratio of the data.
In particular, we are interested in two aspects:
how the probability distributions are modified, and
how much scatter in a given \iell\ value can occur,
when considering maps with noise integrated from one
year to four years and then to a coadded map.
We have performed a large number of simulations ($\sim$ 20000) 
to assess this dependence. In particular, we simulate a noiseless 
CMB sky, then add noise corresponding to the 53A channel for 1-year,
add a second year of noise to create an effective 2-year map,
then similarly to create the equivalent of a 4-year sky,
a 4-year 53 GHz sum map, and finally a 4-year coadded 53 and 90
GHz map. In this way, we specifically account for the effect 
of noise integration over the mission and its impact on the
underlying bispectrum signal. 

Since we have been primarily concerned with the \ing\
amplitude, we shall concentrate on this mode here.
Interestingly, the 6 bispectrum probability distributions 
for the cases considered -- from noiseless CMB sky to 
a coadded map -- have essentially identical probability
distributions (to the extent that plotting them would be
superfluous). In fact, we have explicitly made use
of this fact above, where we have compared the results from
different maps and time-periods directly in all Figures. Here,
we have shown that this is valid, and that a change in
\ing\ value from one map to another can be simply interpreted
against a single underlying probability function.
However, whilst the statistical distribution for \ing\,
and indeed for all bispectrum amplitudes at least out to 
$\ell$ = 30 as considered in this paper,
is essentially unchanged by varying noise contamination,
it is by no means the case that the bispectrum amplitude
for a single CMB realisation is unaffected by instrument noise.

In fact, it is common that, for a given CMB realisation,
there is some scatter in the determined bispectrum amplitude with
noise integration. To quantify this, we have considered the 
difference between both successive pairs of estimates of \ing\ 
(eg. between a simulated 2-year and 4-year map)
and between the underlying CMB amplitude and successive estimates
of \ing. In all cases, the rms difference is $\sim$ 0.25,
although the distributions of the differences are more peaked
than Gaussian with slightly broader tails.
Fig.~11 summarises our findings. 
A particular value of the \ing\ amplitude for the CMB sky can
be enhanced or suppressed by the addition of noise. 
Nevertheless, it is not common for an initially large
CMB \ing\ amplitude to be obscured by noise, nor a small
CMB amplitude to be enhanced to a level suggestive of non-Gaussianity.
In fact, in only 5\% of the simulated cases do we find that there
is a change in the \ing\ value of over 0.5 between the noiseless CMB
case and the coadded sky map. Furthermore, only $\sim$ 0.7\%
of the simulations have CMB values greater than 0.8 and coadded
amplitudes less than 0.2, or the reverse.




\clearpage

\figcaption{The probability distribution functions
for the \iell\ statistic of a Gaussian CMB sky 
and random Gaussian noise appropriate to the coadded 53 and 90 GHz
sky maps. Dashed line: the observed value of \iell\ for the coadded
sky map in Galactic coordinates.}

\figcaption{Upper: the coadded 53 and 90 GHz sky map in 12 
different coordinate systems. The middle column corresponds to 4
reference frames of astronomical significance, the remaining 8 
correspond to rotations of the sky designed to place the Galactic
plane with specific geometries with respect to the underlying
quadcube pixelisation scheme. Lower: the unfolded-T projection of the
quadcube. The right-hand figure shows the rotated orientation of the
Galactic plane relative to the left-hand figure.}

\figcaption{Variation of the \iell\ statistic computed with the
coadded 53 and 90 GHz data for 12 coordinate systems.
These include 4 astronomically important reference frames --
Galactic (G), Ecliptic (E), Celestial (C) and Supergalactic (Z) --
plus 8 others designed to place the Galactic plane in specific
orientations with respect to the \cobe\ quadcube.}

\figcaption{The \ing\ bispectrum amplitude for the 53A, 53B, 90A,
90B and their sum (S) and difference (D) maps as a function of our
12 coordinate systems.}

\figcaption{The 53 and 90 GHz \ing\ bispectrum amplitudes
for the yearly and two-yearly Galactic sky maps. The index scheme is such that
0010 corresponds to the second year of data, 1100 to the third plus
fourth years of data.}

\figcaption{Results of the jack-knife analysis applied to the
Galactic, Ecliptic and System 1 coadded sky maps. Note that the
exclusion of various pixel groupings can both enhance and suppress
the \ing\ bispectrum amplitude. The maps of the bispectrum amplitudes
have had the mean value subtracted.}

\figcaption{Top five panels: selected \iell\ amplitudes for the 31 GHz 
A,B sum and difference maps in 4 coordinate systems. 
Key: channel A = $\triangle$; channel B = $\Diamond$; 
sum map = $+$; difference map = $\times$.
Bottom panel: \iell\ for the 31A channel both excluding (open circles) 
and including (filled circles)
the eclipse data. Note the marked increase at $\ell$ = 16 when the
eclipse data is included.} 

\clearpage

\figcaption{The temperature difference of the coadded sky maps in
Galactic coordinates including and excluding the eclipse data.}

\figcaption{The \iell\ statistic for the coadded, 53 and 90 GHz
A,B and sum (S) sky maps 
both including (filled circles) and excluding (open circles)
the eclipse data.}

\figcaption{Results of fitting a simple power-law cosmological
model, parameterised by a quadrupole normalisation \qrms\ and
a spectral slope \n, to the coadded 53 and 90 GHz Galactic
sky maps. The dashed line corresponds to the fit 
including the eclipse data, the solid line excludes 
this period. The best-fit values are represented by the
open circle when the eclipse period is included and the solid circle
excluding this data. The shift in best-fit values is
comparable to that seen between Galactic and Ecliptic sky maps.}

\figcaption{Simulations of the \ing\ bispectrum amplitude 
for an initial Gaussian CMB sky (\lq CMB') to which appropriate Gaussian
noise levels are successively added to replicate the signal-to-noise
properties of a 1-year 53A sky map (\lq 0001'),
a 2-year 53A sky map (\lq 0011'), a 4-year 53A sky map (\lq 1111'),
a 53 GHz sum map (\lq Sum') and finally a coadded 53 plus 90 GHz
sky map (\lq Coadd'). The top panel represents the worst case of
differences between successive \lq maps' for an initial CMB sky
of large \ing\ value. The lowest panel shows the worst case 
when one is left with a coadded sky of large \ing. Some more
typical variations are shown in between.}

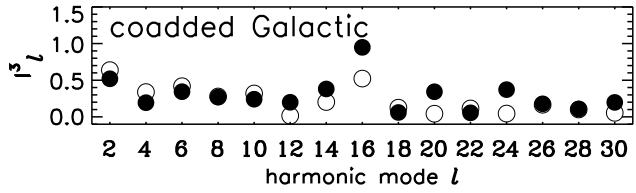
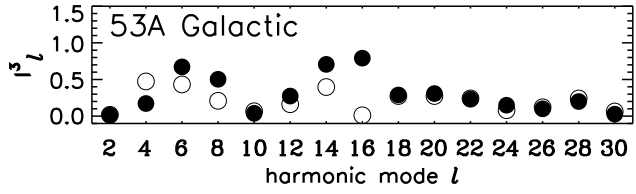
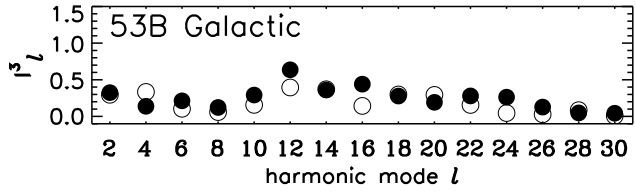
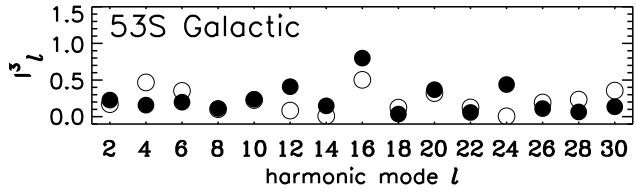
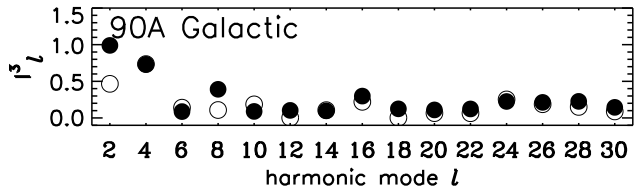
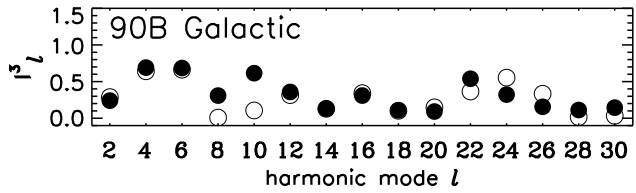
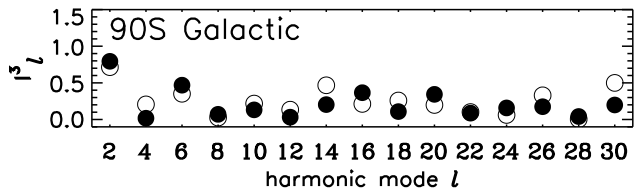
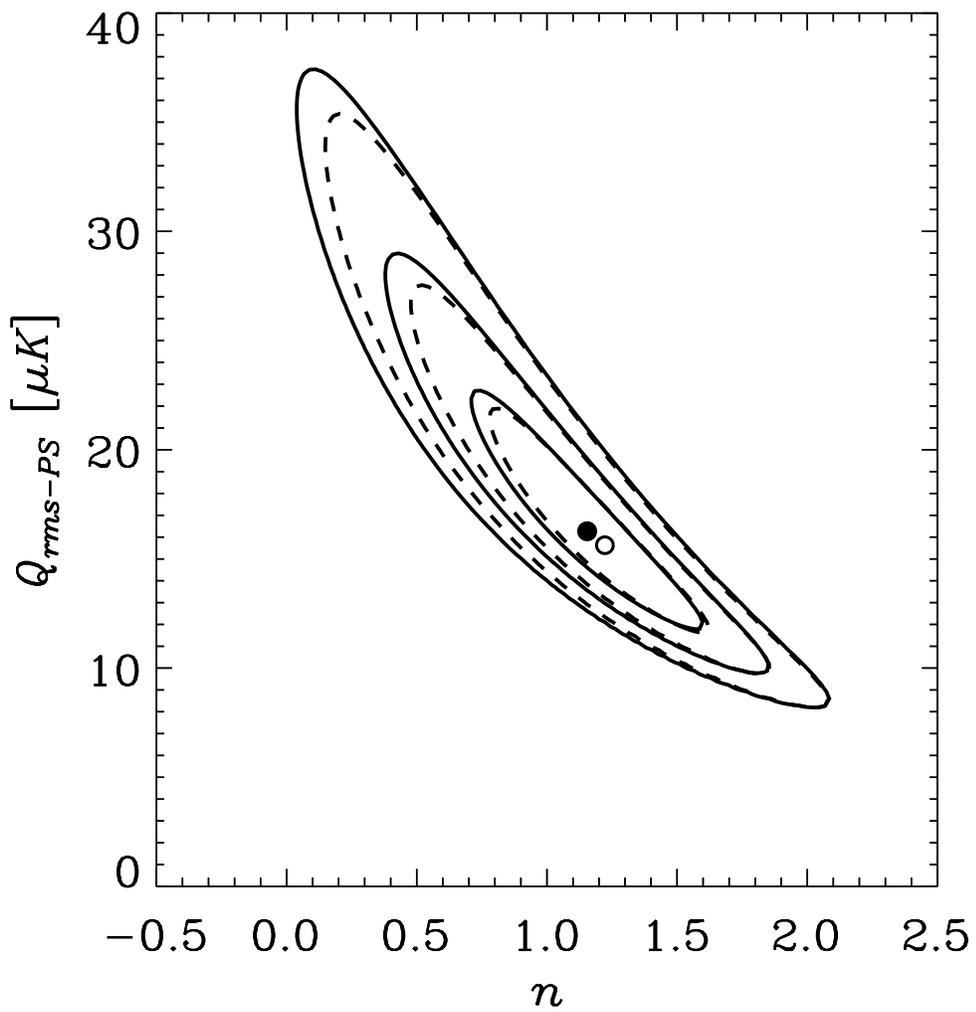
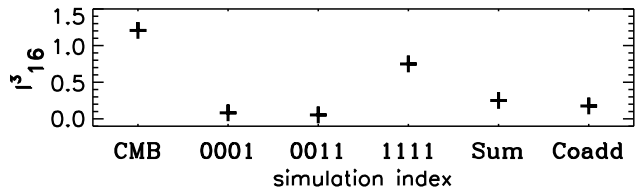
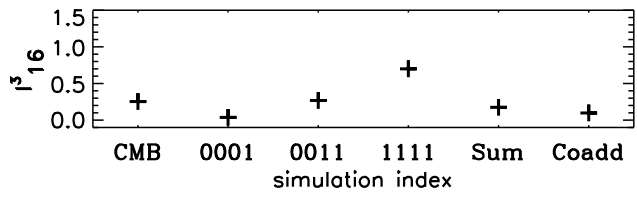
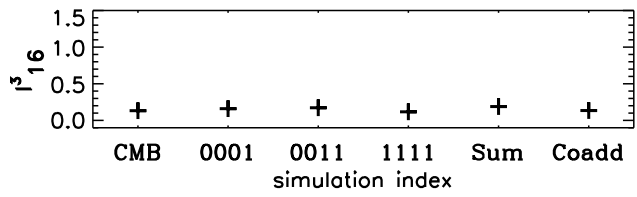
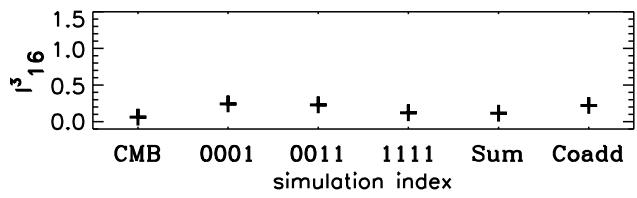
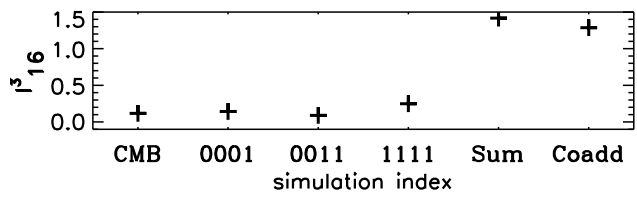
\end{document}